\documentstyle[aps,psfig]{revtex}

\def\be{\begin{equation}}
\def\ee{\end{equation}}

\begin{document}

\title{Damping time and stability of density fermion perturbations in the expanding universe}
\author{Costantino Sigismondi}
\address{Yale University, Dept. of Astronomy, 260 Whitney Avenue 06551 New Haven, CT USA; Osservatorio Astronomico di Roma, V.le del Parco Mellini 84 Roma, Italy and  I.C.R.A.-International Center for Relativistic Astrophysics-
University of Rome``La Sapienza", Physics Department, P.le A. Moro 5, 00185 Rome, Italy}
\author{Simonetta Filippi}
\address{I.C.R.A. and
University CBM, Via E. Longoni 83, 00155 Rome, Italy}
\author{Luis Alberto S\'anchez}
\address{Universidad Nacional de Colombia, A.A. 3840, Medellin,Colombia}
\author{Remo Ruffini}
\address{I.C.R.A.-International Center for Relativistic Astrophysics-,
University of Rome``La Sapienza", Physics Department, P.le A. Moro 5, 00185 Rome, Italy}

\maketitle

%\abstracts{
\abstract{
The classic problem of the growth of density perturbations in an expanding
Newtonian universe
is revisited following the work of Bisnovatyi-Kogan and Zel'dovich. We propose a more general analytical
approach: a system of free particles
satisfying semi-degenerate Fermi-Dirac statistics on the background
of an exact expanding
solution is examined in the linear approximation. 
This differs from the corresponding work of Bisnovatyi-Kogan and Zel'dovich 
where classical particles fulfilling Maxwell-Boltzmann statistics were considered.
The solutions of the
Boltzmann equation are obtained
by the method of characteristics.  An expression for the
damping time of a decaying
solution is discussed and a zone in which free streaming is hampered is found, corresponding to wavelengths less than the Jeans one.
In the evolution of the system, due to the decrease of the Jeans length, those perturbations may lead to gravitational collapse.
At variance with current opinions,
we deduce that perturbations with 
$\lambda \geq \lambda_{J~Max}/1.48$ are able to 
% ``onset" is a noun, not a verb 
generate structures and 
the lower limit for substructures mass is 
$M =M_{J~~max}/(1.48)^3~\approx M_{J~~max}/3$, where $M_{J~~max}$ is the maximum value of the Jeans mass.}

%\vglue 1cm

\section*{PACS}

98.80.Hw Mathematical and relativistic aspects of cosmology

05.30.Fk Fermion systems

05.70.Jk Critical point phenomena

95.35.+d Dark matter

\bigskip
\section*{Introduction}

Bisnovatyi-Kogan and Zel'dovich (1970)\cite{BKZ} approached analytically the study of  gravitational clustering of classical particles within the framework of a Newtonian universe. The general relativistic approach is necessary only for perturbations with wavelengths $\lambda \ge \lambda_{Horizon}$ (see e.g., Peebles (1993) \cite{Pee}).
They considered the evolution of density perturbations for particles obeying Maxwell-Boltzmann statistics. They found that for perturbation lengths $\lambda \gg \lambda_{J}$ the regime is the hydrodynamical one, while for $\lambda \ll \lambda_{J}$, they  found a damping of density fluctuations instead
of the traditional oscillatory behaviour for perturbations in a fluid (see e.g., Binney and Tremaine (1987) \cite{Binney}): they pointed out that short wavelengths perturbations are damped due to some purely kinetic mechanism. 
They studied only small and very large wavelengths with respect to the Jeans length and treated the Jeans length of the perturbation as the limiting length scale between the stable perturbations and the unstable ones.
The cosmological large-scale structure formation scenario in the work of Bisnovatyi-Kogan and Zeldovich (devoted to the study of a typical astrophysical collisionless system, like the edge of stellar clusters or clusters of galaxies) was the framework in which the self-gravity of collisionless particles (stars or galaxies) is in competition with the  universal expansion to create a condensation of matter.
This approach was not taken in order to study a 
{\sl top-down} scenario in which big perturbations have successive fragmentations, even if the Zel'dovich {\sl top-down} scenario of {\sl pancakes} was proposed during those years \cite{ZSU}. 

On the other hand Gilbert (1966) \cite{Gil}, numerically solving an integral equation, has claimed that collisionless damping is similar to the Landau damping in plasma physics. Recall that in  plasma physics the Landau damping is a phenomenon in which the energy of a  definite wavelength perturbation is efficiently dissipated by collective motions of the system \cite{Book}.

Bisnovatyi-Kogan and Novikov (1980) \cite{BKN} and Bond, Efstathiou and Silk (1980) \cite{Bon} have discussed the physics of neutrino masses in the expanding universe. When massive neutrinos are relativistic their Jeans mass $M_J$ is larger than the horizon mass $M_H$, while when they become nonrelativistic the corresponding Jeans mass becomes smaller than the horizon mass, making possible collapse of masses $M_H \ge M \ge M_J$ due to gravitational instability. They obtain the analytic formula for the red shift $z_{n.r.}(m_{\nu}) \propto {m_{\nu}}^2$ at which the Jeans mass of the neutrinos enters the horizon, at this moment the whole Jeans mass is causally connected and its gravitational collapse starts (see fig. 2).
The relation between mass and the dimensions of large scale structure can be recovered: scaling to the present radius of the universe $\lambda_{J~~n.r.}(m_{\nu}) \approx 120 Mpc (\frac{m_{\nu}}{10 eV})^{-1}$. The neutrino is the first candidate for the dark matter, and leads naturally to a {\sl top-down} scenario, owing to its low mass value.  

The work of Bond and Szalay (1983) \cite{BSZ}, based on a numerical approach, assessed the damping of the perturbations smaller than the Jeans length. They have followed a general relativistic approach based on the Gilbert (1966) method \cite{Gil}. This method contains the first order approximation of the equations of motion as in the Bisnovatyi-Kogan and Zel'dovich one. 

White, Davis and Frenk (1984) \cite{Whi} have shown that free streaming should only permit the onset of very large and hot clusters of gas (using the fraction of baryonic and therefore visible matter as a one-to-one tracer for the dark matter distribution) at $KT \approx 20 KeV$ and higher, too luminous in X-rays with respect to the observed ones at this epoch (the hottest known is the Perseus cluster at $KT=7 KeV$).

The alternative cosmological model  to a neutrino-dominated universe appeared in 1984 with the introduction of Cold Dark Matter by Primack and Blumenthal \cite{Pri}, and the Hot Dark Matter models (including neutrinos) began to be somewhat neglected \cite{Tur}.

We have used the method of Bisnovatyi-Kogan and Zel'dovich in a detailed analytic study of the fragmentation for gravitational instability due to the Jeans mass of fermions, extending their approach with care to wavelengths on the order of the Jeans length. We find a more general expression for the damping time formula.

We have evaluated the role of free streaming by studying the solution of the  gravitational Poisson equation for a fermion density perturbation in the expanding universe. We have obtained a new lower limit for the free streaming length that is 1.48 times smaller than the Jeans length.
The fundamental role for the process of structure formation appears to be linked to this new length, determined by the dynamics of the 
self-gravitating fermion system, and not to the traditional free-streaming length determined only by the kinematics of the particles.

In section I, we present the mathematical tools for treating the growth of fermion density perturbations in the expanding universe. In section II, we review the free-streaming effect, while in section III, we define the damping criteria applied to the decaying solution of section I. In section IV, the properties of the nondecaying solutions belonging to the {\sl instability shell} are discussed. Finally in section V, we outline the conclusions.
The appendix applies the method of characteristics to integrate the Boltzmann equation.
 
\section{Solution of the linearized Boltzmann equation}

A kinetic treatment is necessary to analyze  the stability of a system
of weakly interacting massive 
particles after their decoupling when the time between collisions is much greater than the
characteristic hydrodynamic time (i.e., the crossing time $t_c = \frac{l}{v_s}$, where $l$ is the
linear dimension of the system and $v_s$ is the velocity of sound {\sl or, better, the average sound velocity}).
We consider a self-gravitating system of particles in the background of an exact expanding solution and we examine its 
stability in the linear approximation.

We describe the collisionless matter with a distribution
function $f(\vec{x}, \vec{v}, t)$ satisfying
the kinetic Boltzmann equation
\be
\frac{\partial f}{\partial t}+\vec{v}\frac{\partial f}{\partial \vec{x}}
- \nabla \Phi \frac{\partial f}{\partial \vec{v}}=0 ,
\label{bol}
\ee
and the Poisson equation
\be
\nabla^{2}\Phi=4\pi G\rho ,
\ee
where
$G$ is the gravitational constant and the total density at $ \vec{x}$  is
\be
\rho(\vec{x})=\int f(\vec{x}, \vec{v}, t)\,d\vec{v}.
\ee

We study the development of perturbations in a Newtonian universe 
with critical density
containing collisionless particles.

%equazioni introduttive!!!
It is known that for such an universe
\be
\rho_{0} = \frac{1}{6\pi G t^2} \,\,  ; \,\, \Phi_{0} = \frac{2}{3} \pi G
\rho_0 {x_i}^2
\ee
(see e.g. Peebles (1993) \cite{Pee}).
Solving the Boltzmann equation by the method of characteristics (Appendix), 
we find that the comoving velocity is an integral of the motion, so that the solution
$f_0$ of the unperturbed Boltzmann equation for the distribution function is an arbitrary function of the comoving 
velocity $u_i = v_i t^{\frac{2}{3}}-{\frac{2}{3}}x_i t^{-{\frac{1}{3}}}$:
\be
f_0 = \alpha f(\vec{u}).
\ee
The constant $\alpha$ is determined by imposing the
condition of self-consistency \cite{BKZ}:
\be
 \int f_0 d^3 v = \frac{1}{6\pi G t^2} .
\ee

The distribution function describing the collisionless matter is 
$f(\vec{x}, \vec{v}, t)$ such that $f=f_0+f_1$  and the
gravitational potential is $\Phi=\Phi_0+\Phi_1$ 
($f_1$ and $\Phi_1$ are small perturbations) satisfying
the kinetic Boltzmann equation.
Following Bisnovatyi-Kogan and Zel'dovich \cite{BKZ}
we seek the solutions in the form
\be
\label{solpr}
\Phi_1 = e^{i\vec{k} \cdot \vec{\xi}} \phi(t)  \; \, ,  \; \,
f_1 = e^{i\vec{k} \cdot \vec{\xi}} f(t) ,
\ee
where $\xi = x/t^{\frac{2}{3}}$ is the comoving coordinate and $\vec{k}=(k,0,0)$.
The solution of the linearized kinetic Boltzmann equation is
\be
\label{solpr1}
f_1 = \int_0^t \frac{\partial \Phi_1}{\partial \vec{x'}}
\frac{\partial f_0}{\partial \vec{v'}} dt' ,\, f_1(0) = 0 .
\label{f1}
\ee

Substituing the expression (\ref{solpr}) into (\ref{solpr1}), using the relations (\ref{x'}), (\ref{v'}) for
$x'_i$ and $v'_i$ in the Appendix, and integrating by parts, we  have
\be
f_1 = i k \frac{\partial f_0}{\partial u_1}
\int_0^t  \phi(t') e^{3 i k u_1 [t^{-{\frac{1}{3}}}- (t')^{-{\frac{1}{3}}}]
}
 dt'.
\label{f}
\ee
Here $u_1$ is the component of the comoving velocity $\vec{u}$ along the direction of the comoving wave number $\vec{k}$.
The Poisson equation, using the expression (\ref{solpr}) becomes
\be
- \frac{k^2 \phi(t)}{t^{\frac{4}{3}}} = 4\pi G \int f d^3 v .
\label{pt}
\ee
Then integrating Equation (\ref{f}) over the velocities and substituting the
result into eq.(\ref{pt}), we obtain the following integral equation for $\phi$:
\be
t^{\frac{2}{3}} {\phi}(t) + \frac{16 \pi^2 G}{k}
\int_0^t \phi(t') dt'\int_0^\infty du~ u f_0 \sin(3ku\tau)=0 ,
\label{eqp}
\ee
where $u=u_1$ is the comoving velocity,
$\tau=t^{-\frac{1}{3}}-t'^{-\frac{1}{3}}$,
and $k$ is the comoving wave number.
We are considering the universe with critical density filled by fermions, decoupling in the relativistic regime, so that the unperturbed 
semi-degenerate Fermi-Dirac distribution function $f_0$ is
\be
f_0(\vec{u}) = \frac{9}{8\pi^2 G <u>^3 g(\xi)}  
\displaystyle{\frac {1}{e^{\frac{3 u }{<u>} - \xi} + 1}}
\label{f0}
\ee
where  
\be
<u>=\frac{3k_B T_{\nu}(t)}{m c},
\label{velocity}
\ee
and
$\xi=\frac{\mu}{k_B T}$ is the degeneracy parameter, 
$k_B$ is the Boltzmann constant and $\mu$ is the chemical potential.
The normalization function is $g(\xi)=\frac{1}{3}\xi^3+2\zeta_{R}(2)\xi+2\sum_{s=1}^{\infty}
\frac{(-1)^{s+1}}{s^3}e^{-s\xi}$, where $~\zeta_{R}(n)=\sum_{s=1}^{\infty}
\frac{1}{s^n}~$ is the Riemann 
function of index $n$.
In equation (\ref{velocity}) the relation between energy (here at equipartition) and momentum for those fermions is in the relativistic regime $E=cp$ because we assume that they are decoupled from the radiation and the matter while relativistic. 
The fact that the particles are relativistic at decoupling only depends on
the weak interactions and not on the fermion mass \cite{Oha}. For the Liouville theorem  extended to the expanding universe \cite{Ehlers}, this relation holds at all times.
In formula (\ref{f0}) the comoving velocity $u=u_1$ along one direction is multiplied by three under the hypothesis of spatial isotropy.

Inserting the distribution function (\ref{f0}) in the integral equation (\ref{eqp}),
we obtain
\be
t^{\frac{2}{3}} {\phi}(t) + \frac{2}{\alpha g(\xi)}
\int_0^t \phi(t') dt'\int_0^\infty dy \frac{y \sin(\alpha y \tau)}
{e^{y-\xi}+1}=0
\label{eqp1}
\ee
where $\alpha=k<u>$, $y=\frac{3u}{<u>}$ and
$<u>=<v^2>^{\frac{1}{2}}t^{\frac{2}{3}}$.
Defining the Jeans length as
\be
\lambda_J=<v^2>^{\frac{1}{2}}(\frac{\pi}{G \rho})^{\frac{1}{2}}
\ee
at a given time $t=t_0$, using also $\rho=\frac{1}{6 \pi G t^2}$,
we can write
\be
\alpha=k<u>=\frac{2}{\sqrt{6}}(\frac{k}{k_J}) t_0^{\frac{1}{3}}=
F\bar{\alpha} t_0^{\frac{1}{3}}
\ee
where $\bar{\alpha}=\frac{k}{k_J}$, and $F=\frac{2}{\sqrt{6}}\approx1.18$.
Defining 
\be
x=F (\frac{t_0}{t})^{\frac{1}{3}}
\label{xf}
\ee
 and
\be
x'=F (\frac{t_0}{t'})^{\frac{1}{3}}
\label{x'f}
\ee
 and $\eta=x'-x$,
 the dimensionless integral equation is
\be
\phi(x) - \frac{6 x^2}{\bar{\alpha} g(\xi)}
\int_0^\infty d\eta \frac{ \phi(x+\eta)}{(x+\eta)^4}
\Gamma(\bar{\alpha},\xi,\eta)=0
\label{eqint1}
\ee
with
\be
\Gamma(\bar{\alpha},\xi,\eta)=
\int_0^\infty dy \frac{y \sin(\bar{\alpha}\eta y)}
{e^{y-\xi}+1} .
\label{eqint2}
\ee

A numerical analysis shows that for fixed $ \bar{\alpha}$ and $\xi$, the function $\Gamma$ increases and
decreases rapidly with $\eta$, having a maximum at $\eta_0$.

The integral in the equation (\ref{eqint1}) can be calculated by the method
of steepest descents \cite{Arf}.
We approximate
$
\Gamma(\bar{\alpha},\xi,\eta)\approx e^{\bar{\alpha} h(\eta,\xi)}
$
so that
\be
\int_0^\infty d\eta \frac{ \phi(x+\eta)}{(x+\eta)^4}e^{\bar{\alpha}
h(\eta,\xi)}
= {\frac{\sqrt{2\pi} \phi(x+\eta_0) e^{\bar{\alpha}
h(\eta_0,\xi)}}{(x+\eta_0)^4 |\bar{\alpha}
h"(\eta_0,\xi)|^{\frac{1}{2}}}} .
\ee

It is important to note that the function $\Gamma(\bar{\alpha},\xi,\eta)$ is not a true function of three variables, but only a function of two variables: $\Gamma(\bar{\alpha}\eta,\xi)$. Thus with the variation of the parameter $\bar{\alpha}$, the $\Gamma$ function undergoes  a conformal transformation along the $\eta$ axis: a contraction as long as $\bar{\alpha}$ grows, and a dilation if it decays. The maximum of the $\Gamma$ function at $\eta_0$ for a given value of $\bar{\alpha}$ is equivalent in the two-variable framework to a unique maximum at $\eta^{*}=\bar{\alpha}\eta_0=const$ (see fig. 1).
So in the approximation of the $\Gamma$ function by an exponential function in the steepest descent method, there are not distinct constraints for small or large values of $\bar{\alpha}$, while Bisnovatyi-Kogan and Zel'dovich examine these two regimes separately.
 
The integral equation (\ref{eqint1}) becomes
\be
\phi(x) - \frac{6 \sqrt{2\pi} x^2}{\bar{\alpha} g(\xi)}
\frac{ \phi(x+\eta_0)}{(x+\eta_0)^4} {\frac{e^{\bar{\alpha}
h(\eta_0,\xi)}}{|\bar{\alpha}
h"(\eta_0,\xi)|^{\frac{1}{2}}}=0} ,
\label{eqint3}
\ee
with
\be
\bar{\alpha} h''(\eta_0)=-\bar{\alpha}^2 \frac{\int_0^\infty dy \frac{y^3
\sin(\bar{\alpha}\eta_0 y)}
{e^{y-\xi}+1}}{\int_0^\infty dy \frac{y \sin(\bar{\alpha}\eta_0 y)}
{e^{y-\xi}+1}}
\label{eqint4}
\ee
since $\bar{\alpha}h(\eta,\xi)=\ln \Gamma(\bar{\alpha},\eta,\xi).$
In this approximation, as in the Bisnovatyi-Kogan and Zel'dovich work, we can Taylor expand the function $\phi(x+\eta_0)$
and
$ (x+\eta_0)^{-4}$ at $\eta_0=0$, retaining the terms of the first order
in $\eta_0$, leading finally to
the differential equation
\be
\phi(x) - \frac{6 }{\bar{\alpha}^2 g(\xi)}
\frac{ B}{x^2}\eta_0 \phi'(x)=0 ,
\label{eqdiff}
\ee
where 
\be
B=\sqrt{2\pi}\frac{(\int_0^\infty dy \frac{y
\sin(\bar{\alpha}\eta_0 y)}
{e^{y-\xi}+1})^{\frac{3}{2}}}{(\int_0^\infty dy \frac{y^3
\sin(\bar{\alpha}\eta_0 y)}
{e^{y-\xi}+1})^{\frac{1}{2}}} ,
\label{B}
\ee
with the solution
\be
\phi(x)=C \exp(\frac{\bar{\alpha}^2 g(\xi)}{18 B\eta_0}x^3)\label{sol}
\ee
or in terms of time $t$, using (\ref{xf})
\be
\phi(t)=C \exp(\frac{\gamma}{t}) ,
\label{fit}
\ee
and
\be
\gamma=\frac{\bar{\alpha}^2 g(\xi)}{18 B\eta_0}F^3 t_0 .
\label{gamma}
\ee

\section{The free streaming effect}

We want to study whether the density perturbations smaller than the Jeans length will lead to substructures lighter than the Jeans mass, or will be dissipated by the collisionless damping.

The free streaming kinematic scale $\lambda_{fs}$ (see fig. 2)
is the scale that a particle with some mean
velocity has traveled up to time $t$: 
\be
\label{fs}
\lambda_{fs}=\int_0^t dt' a(t')v(t')
\ee
(where the scale factor of the universe $a(t') \propto (z+1)^{-1}$). 

The horizon size
$\lambda_{Horizon}$ is the distance traveled by a photon emitted at $t=0$(see fig. 2). 

The Jeans scale $\lambda_J$ contains a mass of particles whose self gravity is greater than the kinetic pressure (see fig. 2).

In the extremely relativistic regime for particles (see the left side of fig. 2) on scales larger than $\lambda_{fs}\approx \lambda_{Horizon}$,
the
perturbations can neither decay nor grow \cite{Padmanaban}, while
on scales $\lambda_2$ smaller than $\lambda_{fs}$
and smaller than the Jeans mass scale
$\lambda_{J}$, the perturbation decays quickly due
to directional dispersion.  As soon as the particles
become nonrelativistic (see the right side of fig. 2), the perturbations should decay slowly due to velocity dispersion.

When the Jeans length becomes smaller than the horizon length, the perturbations greater in scale than $\lambda_{J}$ and $\lambda_{fs}$ increase (see $\lambda_1$ on the right side of fig. 2 when it crosses the Jeans length curve).

Perturbations started after $z_{n.r.}$ with wavelengths $\lambda_3$ that are greater than $\lambda_{fs}$ (calculated integrating \ref{fs} from $t_O > t_{n.r.}$) will survive and collapse when crossing the $\lambda_J$ curve, while the ones like $\lambda_4$ will be dissipated by the free streaming.  

The current opinion \cite{Whi,Dur} is that the free streaming scale is the lower limit for surviving perturbations that may lead to subsequent gravitational collapse. Below this scale, it is generally assumed that all perturbations are dissipated. In what follows we present an alternative approach to this problem.

\section{Damping criteria}

The characteristic time for damping starting from the time $t_0$ is the time
interval $T$
in which the density perturbation is reduced to $\frac{1}{e}$ of its value at the time
$t_0$.
From the expression (\ref{fit}) one finds
\be
T=\frac{t_0}{\gamma'-1} ,
\label{tdam}
\ee
with $\gamma'=\frac{\gamma}{t_0}$ and
\be
\gamma'=\frac{g(\xi)F^3}{18 B(\xi)\eta_0}(\frac{k}{k_J})^2 .
\ee
If the parameter $\gamma'$ is near 1, the damping time can be very long
and the
density perturbations of the corresponding wave number $k$  survive the
collisionless damping
process.

The expression (\ref{tdam}) is more general than the characteristic time of damping considered by Bisnovatyi-Kogan and Zel'dovich (1970) (equation 20).

Recalling that $\bar{\alpha}\eta_0=\eta^{*}$ ($\eta_0$ makes the numerator of $B$ in (\ref{B}) a maximum  for a given value of
$\bar{\alpha}$) and that $\bar{\alpha}=\frac{k}{k_J}$, where $k_J$ is the comoving wave
number corresponding to the Jeans length \cite{RSO}, we can derive the lowest value of the wave number $k$ for which the equation (\ref{tdam}) has negative values, corresponding to undamped solutions
\be
\frac{k}{k_J}~~~=~~~^3\sqrt{\frac{18 B(\xi)\eta^{*}}{g(\xi)F^3}}\approx 1.25~~~=~~~\bar{\alpha_1} .
\label{klim}
\ee

In the expression (\ref{klim}) $F\approx1.18$, $g(\xi)\approx1.80$ for $\xi=0$,
$B \approx 0.60 $ and $\eta^{*} \approx 0.54$,  we have that fluctuations with 
$\bar{\alpha_1} \leq 1.25$ are not damped.

Another criterion defining the 
undamped perturbations is evident from fig. 3, in which we have represented
the
temporal evolution of the relative amplitude of the fermionic density
perturbations $\delta\over{\delta_{0}}$, where $\delta \propto a(t)\phi (t)$ (see e.g. Peebles (1993)  \cite{Pee}) and $\delta=\frac{(\delta\rho)_0}{\rho_0}$ is the density contrast at time $t_0$. 

Setting $\xi=0$ leads to
\be
\frac{\delta}{\delta_{0}}~~=~~
(\frac{a}{a_0}) \exp{[0.51~\bar{\alpha}^3((\frac{a}{a_0})^{-\frac{3}{2}}-1)]} .
\ee

For each curve we have calculated the abscissa ($\frac{a}{a_0}=(\frac{t}{t_0})^{\frac{2}{3}}$, the expansion parameter of the universe) of the minimum
point  

\be
\left(\frac{a}{a_0}\right)_{min} ~\approx ~  ~ 0.83  ~  ~ {\bar{\alpha}}^2.
\ee

For a curve attaining the value 
\be
{\delta(\left(\frac{a}{a_0}\right)_{min})\over{\delta_{0}}} = \frac{1}{e} ,  
\label{tras}
\ee
we assume that the perturbation is undergoing a significant damping; numerically solving  
the trascendental equation (\ref{tras}) we  obtain the value 
\be
\frac{k}{k_J}~~~ \approx~~~ 1.71~~~=~~~\bar{\alpha_2};
\ee
from this second criterion the fluctuations with 
$\bar{\alpha_2} \leq 1.71$ are not damped.

Summarizing the results found using  these two criteria:
 
A perturbation in the potential or in the density contrast function decays exponentially with time for short wavelengths.

Furthermore in a given range of wavelengths near the classical Jeans length,
this decaying slows radically and the perturbations stabilize.

Due to the weak self-gravity the damping of the perturbations is faster for small linear
dimensions (wavelengths). This behaviour is shown in the fig. 3 for high values of $\bar{\alpha}$.

The physical mechanism responsible for the damping of the perturbations, in
our model, is the free streaming effect.
It is to be intended as a kinematical effect, whose amount depends on the statistics of the particles.
 
Self-gravity is the competing mechanism for the growth of perturbations, which is a collective effect.

The dynamical free-streaming scale results from these competing mechanisms and it also
defines  a length that is the lower limit for density perturbations capable of forming structures.

\section{The properties of the {\sl instability shell} and physical remarks about the solutions}

In figure 4 we describe the processes of growth, stability and decay of fermion density perturbations.
The figure is divided into two parts using a vertical line that corresponds to the transition from the ultra-relativistic to the non-relativistic regime.

On the left part of the figure we consider perturbations in the ultra relativistic regime. We select a perturbation of comoving length $\lambda_2$ which evolves along an horizontal line: for $\lambda_2 > \lambda_{horizon}$ the perturbation is stable and has no damping since the different parts of the perturbation are not causally connected. For $\lambda_2 < \lambda_{horizon} < \lambda_{Jeans}$ the self gravity starts to act and the perturbation is quickly damped by free streaming.

On the right part of figure 4 we consider the perturbations in the 
nonrelativistic regime. 

From the two damping criteria we have found that some perturbations of wavelengths smaller than both the Jeans length and the free-streaming length are stable and undamped.
Selecting a perturbation with an intermediate value of $\frac{k}{k_J}=\bar{\alpha}$ between 1.25
(found
with the first criterion) and 1.71 (found with the second), we see that
significant damping occurs only
for $\frac{k}{k_J}=\bar{\alpha} > 1.48$.

This is the definition of the {\sl instability shell}, in which the perturbations with $\bar{\alpha} \leq 1.48$ are not damped and may lead to subsequent collapse due to gravitational instability.

So we can say that the free-streaming equation (\ref{fs}) describes the "kinematical" free-streaming, for which self gravity has no effect.
The crossing of the perturbation length $\lambda_1$ (and $\lambda_3$) with the kinematical free-streaming length curves (that are extended as gray lines into the {\sl instability shell} for showing the effect), in the right-hand region, does not imply the damping of the perturbation, because this length belongs to the {\sl instability shell}.

Our solutions satisfy the Poisson equation requiring a causal connection between particles, so they are not fully extendible to the relativistic regime, for which the Jeans scale is larger than the horizon.

Meanwhile it is reasonable to assume a continuity of our solutions for the {\sl instability shell} towards the relativistic regime.
The {\sl instability shell} becomes the region between the horizon and kinematical free-streaming lengths, and its lower limit goes continously from the kinematical free-streaming length to $\frac{\lambda_{J~Max}}{1.48}$.

Consequently we can deduce that perturbations with $\lambda_1 \geq \lambda_{J~Max}/1.48$ are able to generate structures.
The lower limit for  substructures (with respect to the maximum Jeans length) mass is $M = M_{J~~max}/(1.48)^3$.

Moreover for a starting perturbation of length $\lambda_3$ while the particles are nonrelativistic, the kinematical free streaming is not efficient, $\lambda_3$ is greater than the lower limit of the {\sl instability shell}, and this substructure being stable, it can start the gravitational collapse when $\lambda_J$ becomes equal to $\lambda_3$.
For $\lambda_4 < \lambda_3$ the perturbation is dissipated as in the classical case (see fig. 2).
In the bottom right-hand part of figure 4 (for $\lambda_3$ and $\lambda_4$) one again finds on smaller scales and in the nonrelativistic epoch the same behaviour of the dynamical free-streaming length regulated by the {\sl instability shell} for the primordial fluctuations $\lambda_1$ and $\lambda_2$.

%The amplitude of {\sl Instability Shell} have positive derivative as %chemical potential $\xi$ increases. Going toward fully 
%degenerate statistics ($\xi \rightarrow \infty $) the 
%quantum statistical effects dominate the dynamics and the range 
%allowed for substructures formation (whith respect to a given 
%$\lambda_J$) extends to smaller wavelengths. Going toward 
%classical limit ($\xi \rightarrow -\infty $) the amplitude 
%of {\sl Instability Shell} becomes narrower and substructures 
%smaller than $\lambda_J$ are not expected. 

\section{Conclusions}

The discovery of the {\sl instability shell}, which depends on the statistical properties of the particles involved in the clustering phenomena, has clarified  the role of free-streaming and furthermore leads to a better definition of it.
We have introduced the concept of kinematical free streaming for the traditional treatment, which neglects self gravity and microscopical quantum statistics, while we introduce the new concept of dynamical free-streaming which takes into account these effects.

Collisionless damping is a consequence of a purely kinematical effect (due to directional and velocity dispersion) of free streaming \cite{Sig} which does not act like a collective effect which the comparison with Landau damping (even the gravitational one \cite{Kan}) suggests. Free streaming is efficient only in the first stages of the damping  process; after those initial transient stages the dynamics of the whole perturbation acts in the opposite direction, like a true collective effect. The successive damping inefficiency is a consequence of the microscopic statistics adopted for the particles: at the same temperature the fermions' velocity dispersion is less spread towards high values with respect to classical Maxwellian particles, thus the free streaming is less efficient for fermions in comparison with classical particles, as we have found.

The  dynamical free-streaming length should refer to the {\sl instability shell's} properties.
Indeed, for semidegenerate Fermi-Dirac particles, primordial perturbations with wavelengths $\lambda \geq \frac{\lambda_{J~Max}}{1.48}$ must survive damping process, this inequality determines the upper limit for the dynamical free-streaming length.

We have identified an intermediate phase, with respect to the previous classical frameworks, in which the perturbation (within the horizon) neither grows nor decays until the corresponding Jeans mass is reached; at that point the collapse starts.
This property allows structures smaller than the maximum Jeans mass to survive and makes possible their fragmentation \cite{RSS} into smaller substructures that can detach from the Hubble flow.

Furthermore for an induced clump (for instance,  hot gas from active galactic nuclei explosions or generated during processes of galaxy and cluster formation \cite{Nor}) of lengths $\lambda \geq \lambda_J/1.48$, while the particles are nonrelativistic, the perturbation belongs to the {\sl instability shell}.
It will lead a substructure over a relatively rapid time scale, via the Jeans instability. 

Moreover the critical ratio $\bar{\alpha}=1.48$ so obtained is independent of the scale of the
perturbations considered, so this property is scale invariant and can be valid even after more fragmentations and may even lead to local fractal-like distribution of matter \cite{RST}.

\appendix
\section*{}

The method of characteristics is applied to solve a quasi-linear and homogeneous equation of the form
\be P(x,y,z)\frac{\partial u}{\partial x}+ Q(x,y,z)\frac{\partial u}{\partial y}+
R(x,y,z)\frac{\partial u}{\partial z}=0
\label{ap1}
\ee
in which the functions $P(x,y,z), Q(x,y,z), R(x,y,z)$ are continuous and not  simultaneously vanishing.
Let us consider the continous vector field
$$\vec{F}=P(x,y,z)\vec{i}+ Q(x,y,z)\vec{j}+ R(x,y,z)\vec{k}$$
with $\vec{i}, \vec{j}, \vec{k}$ unit vectors along the coordinate axes. 
The characteristics of (\ref{ap1}) are integral curves of this vector field, defined by integrating the following equations
\be
\frac{dx}{P(x,y,z)}=\frac{dy}{Q(x,y,z)}=\frac{dz}{R(x,y,z)} .
\label{car}
\ee
Two independent integrals of these equations
$\Psi_1 (x,y,z)=C_1$ and $\Psi_2 (x,y,z)=C_2$ represent surfaces whose intersection curves are these characteristics.
The general solution of (\ref{ap1}) can then be expressed in the form
$u=\Phi(C_1,C_2)$.

The characteristics of the Boltzmann equation \cite{BKZ}
\be
\frac{\partial f}{\partial t}+\vec{v} \frac{\partial f}{\partial \vec{x}}-
\frac{\partial \phi}{\partial \vec{x}}\frac{\partial f}{\partial \vec{v}}=0
\label{te}
\ee
are determined by the system:
\be
dt=\frac{dx_i}{v_i}=-\frac{dv_i}{\frac{2}{9}\frac{x_i}{t^2}}
\label{sys}
\ee
from which
we obtain
\be
x_i=\int v_i dt,
\label{x}
\ee
and
\be
\frac{dv_i}{dt}=-\frac{2}{9}\frac{x_i}{t^2} .
\label{vi}
\ee
Substituting eq.(\ref{x}) in eq.(\ref{v}) leads to
$$
\frac{dv_i}{dt}+\frac{2}{9t^2}\int v_i dt=0 ,
$$
and differentiating this with respect to the time $t$ we obtain the second order differential equation
$$
t^2\frac{d^2 v_i}{dt^2}+2t\frac{dv_i}{dt}+\frac{2}{9}v_i=0 .
$$

If $t=e^z$ we have
$$\frac{d^2 v_i}{dz^2}+\frac{dv_i}{dz}+\frac{2}{9}v_i=0 .
$$
The solution is 
\be
v_i=C_{1i} t^{-\frac{1}{3}}-C_{2i} t^{-\frac{2}{3}} .
\label{v}
\ee
Substituting eq. (\ref{v})
in equation (\ref{x}) and integrating, we have
\be
x_i=\frac{3}{2}C_{1i} t^{\frac{2}{3}}-3 C_{2i} t^{\frac{1}{3}} .
\label{xi}
\ee
Combining the two expression (\ref{v}) and (\ref{xi}),
we obtain the six integrals of the system (\ref{sys})
\be
C_{1i}=3v_i t^{\frac{1}{3}}-x_i t^{-\frac{2}{3}} ,
\label{C1}
\ee
\be
C_{2i}=v_i t^{\frac{2}{3}}-\frac{2}{3}x_i t^{-\frac{1}{3}}.
\label{C2}
\ee
An arbitrary function $f(C_{1i},C_{2i})$ of these integrals is a solution of the transport equation (\ref{te}). 

The integral of the motion $C_{2i}$ given by eq. (\ref{C2})
coincides with the comoving velocity $u_i= v_i t^{\frac{2}{3}}-\frac{2}{3}x_i t^{-\frac{1}{3}}$.

Solving the integrals (\ref{C1}) and (\ref{C2}) with respect to $x'$ and $v'$ we have
\be
3v_i t^{\frac{1}{3}}-x_i t^{-\frac{2}{3}}
=3v'_i t^{\frac{1}{3}}-x'_i t^{-\frac{2}{3}}
\ee
\be
v_i t^{\frac{2}{3}}-\frac{2}{3}x_i t^{-\frac{1}{3}}
=v'_i t^{\frac{2}{3}}-\frac{2}{3}x'_i t^{-\frac{1}{3}}
\ee
from which we obtain
\be
x'_i=(3v_i t^{\frac{1}{3}}-x_i t^{-\frac{2}{3}})t'^{\frac{2}{3}}+
(2x_i t^{\frac{-1}{3}}-3v_i t^{\frac{2}{3}})t'^{\frac{1}{3}}
\label{x'}
\ee
and
\be
v'_i=\frac{2}{3}(3v_i t^{\frac{1}{3}}-x_i t^{-\frac{2}{3}})t'^{-\frac{1}{3}}+\frac{1}{3}
(2x_i t^{-\frac{1}{3}}-3v_i t^{\frac{2}{3}})t'^{-\frac{2}{3}} .
\label{v'}
\ee
The equations (\ref{x'}) and (\ref{v'}) are the equations of the trajectories. Both coordinates and velocities of the particle at a given time $t'$ are expressed through coordinates and velocities at time $t$ \cite{BKZ}, \cite{FRI}. These equations are utilized in equation (\ref{f}) making it possible to take into account the cosmological expansion of the background.

\newpage
\vspace{2cm}
\begin{figure}[t]
\centerline{
\hspace{3.5cm}
\psfig{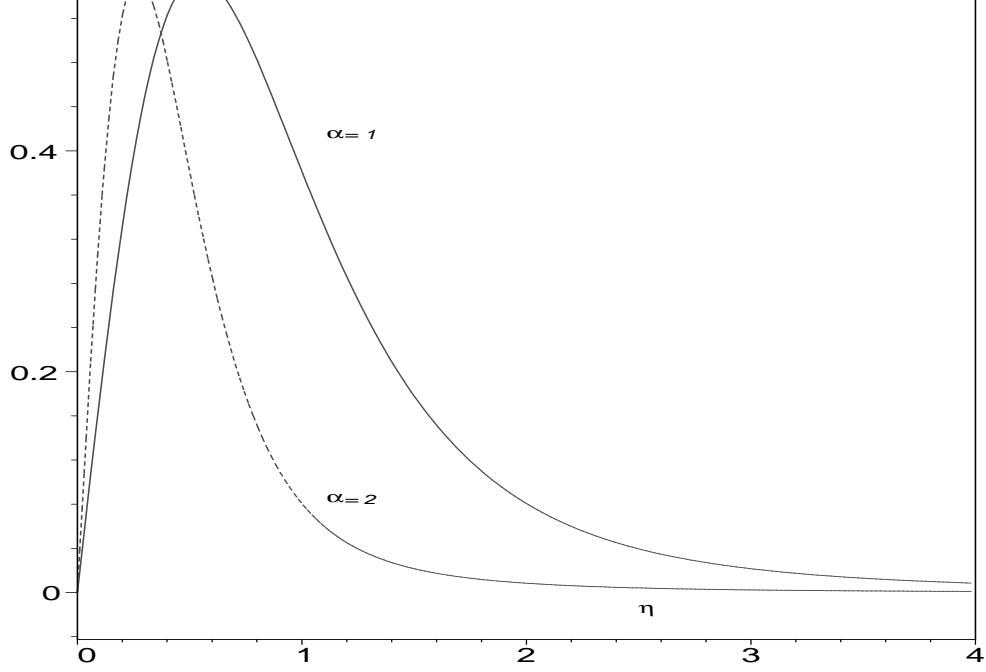}
}
\vspace{2cm}
\caption[]{The function $\Gamma(\eta)$ versus $\eta$ for two values of $\alpha = 1, 2$}
\label{fig: fig 1}
\end{figure}

\newpage

\vspace{2cm}
\begin{figure}[t]
\centerline{
\hspace{4.5cm}
\psfig{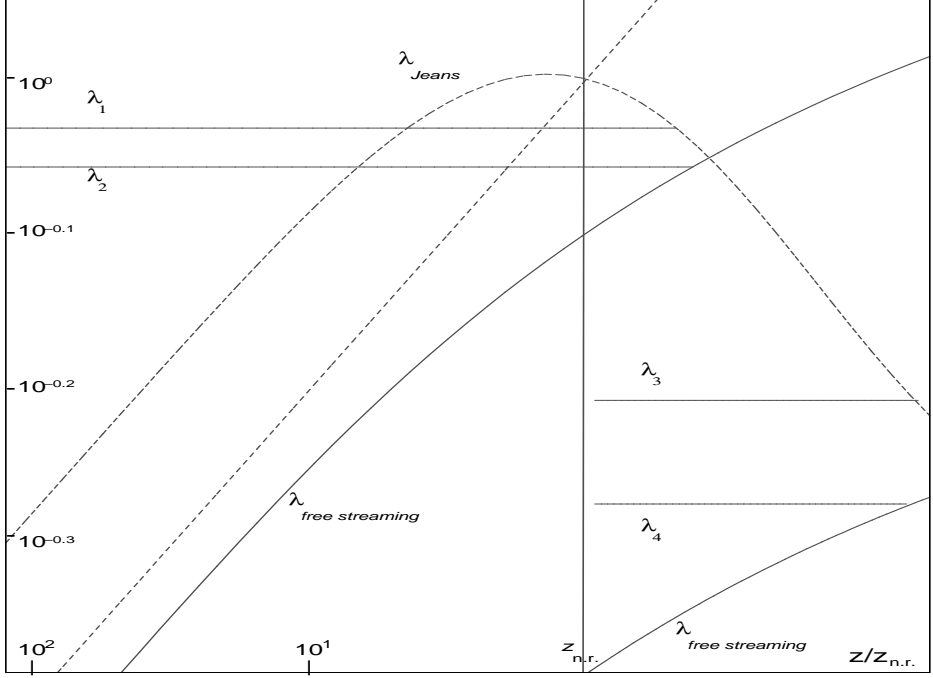}
}
\vspace{2cm}
\caption[]{The traditional free-streaming considerations. The Jeans length (in
dashed-dotted lines); the free-streaming length (in solid lines) and 
the horizon length (in dashed lines) are given as a function of the red shift (qualitative
drawing). 
Perturbations corresponding to a wavelength $\lambda_1$ and $\lambda_3$, by crossing the 
$\lambda_{Jeans}$ line, give origin to a process of collapse, leading to the onset of structures.
The perturbations with wavelength $\lambda_2$ and $\lambda_4$, by crossing the free-streaming 
line, are dissipated. On the lower right side we consider perturbations originating in the 
nonrelativistic regime (see text).
$\lambda_{Jeans max}$ and $z_{n.r.}$ depend on the fermion mass.} 
\label{fig: fig 2}
\end{figure}

\newpage
\vspace{2cm}
\begin{figure}[t]
\centerline{
\psfig{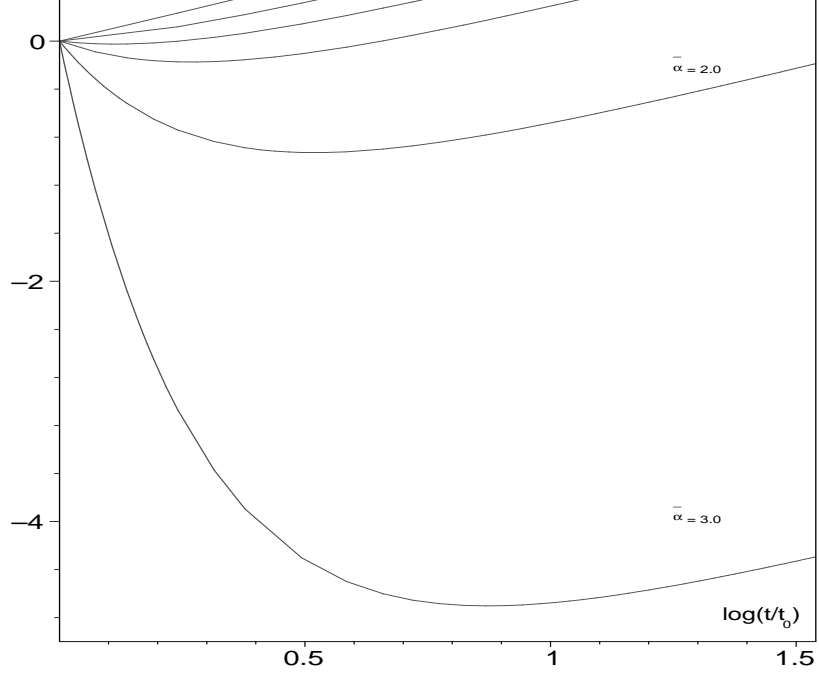}
}
\vspace{2cm}
\caption[]{The normalized density perturbation versus time
for selected values of the ratio $\bar{\alpha}=\frac{k}{k_J}$. Top to bottom:$\bar{\alpha}$ = 0.4,
1.0,
1.255, 1.5, 2.0, 3.0. The axes are in logarithmic scale.}
\label{fig: fig 3}
\end{figure}

\newpage
\vspace{2cm}
\begin{figure} [t]

\centerline{
\hspace{4cm}
\psfig{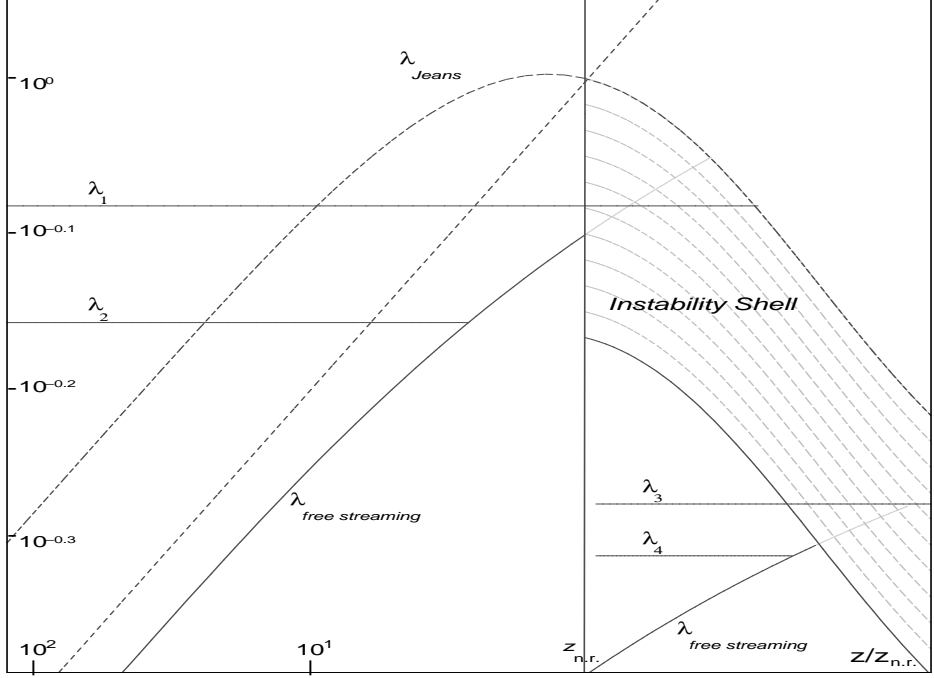}
}
\vspace{2cm}
\caption[]{Properties of the \textsl{instability shell} in the 
nonrelativistic regime. The horizontal lines
corresponding to wavelengths $\lambda_1$ and $\lambda_3$ and attaining the Jeans length curve
(dashed-dotted lines) are the surviving perturbations able to generate
structures, while $\lambda_2$ and $\lambda_4$ are dissipated by the
free-streaming (rising solid lines). At redshift $z \le z_{n.r.}$ (the vertical
line) the \textsl{instability shell} (represented by the region with dashed lines)
is bounded above by the $\lambda_{Jeans}$ and below by the solid line defined
by $\lambda_{inf}=\frac{\lambda_{Jeans}}{1.48}$. The effects of the free streaming are hampered
in the \textsl{instability shell}.
The horizon length is indicated by the dashed lines. On the lower right site we consider perturbations originating in the 
nonrelativistic regime (see text).}

\label{fig: fig 4}
\end{figure}

\end{document}